\pdfoutput=1
\documentclass[journal]{IEEEtran}

\usepackage{cite}

\ifCLASSINFOpdf
  \usepackage[pdftex]{graphicx}

  \DeclareGraphicsExtensions{.pdf,.jpeg,.png}
\else

\fi
\ifCLASSOPTIONcompsoc
\usepackage[caption=false,font=normalsize,labelfon
t=sf,textfont=sf]{subfig}
\else
\usepackage[caption=false,font=footnotesize,subrefformat=parens,labelformat=parens]{subfi
g}
\fi
\usepackage{amsmath}
\usepackage{stix}
\usepackage{threeparttable}
\usepackage{array}
\usepackage{xcolor}
\usepackage{soul}
\soulregister\cite7

\usepackage{pdfpages}
\usepackage{tikz}
\usetikzlibrary{calc,shapes,arrows,chains,arrows.meta}
\usepackage[export]{adjustbox}
\usepackage{upgreek}
\usepackage{makecell}
\usepackage{multirow}
\definecolor{newTextColor}{RGB}{0,0,0} 
\definecolor{modTextColor}{RGB}{0,0,0} 
\definecolor{typoColor}{RGB}{0,0,0}

\newcounter{node}
\usepackage[hidelinks]{hyperref}
\makeatletter
\newcommand{\customlabel}[2]{%
  \protected@write \@auxout {}{\string \newlabel {#1}{{#2}{\thepage}{#2}{#1}{}}}%
  \hypertarget{#1}{#2}
}
\tikzset{autonumbered node/.style={/utils/exec={\stepcounter{node}}},label=above:{\alph{node}}}

\makeatother

\begin{document}

\onecolumn
\pagenumbering{Roman}
\textbf{Author's pre-print}
\newline\newline
$\copyright$ 2024 IEEE. Personal use of this material is permitted. Permission
from IEEE must be obtained for all other uses, in any current or future
media, including reprinting/republishing this material for advertising or
promotional purposes, creating new collective works, for resale or
redistribution to servers or lists, or reuse of any copyrighted
component of this work in other works.

This article has been accepted for publication in IEEE Transactions on Antennas and Propagation. This is the author's version which has not been fully edited and
content may change prior to final publication. 
Citation information: DOI~10.1109/TAP.2024.3407357
\newpage
\twocolumn
\pagenumbering{arabic} 
%
\title{Deriving Characteristic Mode Eigenvalue Behavior Using Subduction of Group Representations}
%
%
%
\author{Lukas~Grundmann, \IEEEmembership{Graduate Student Member, IEEE}, Lukas Warkentin and  Dirk~Manteuffel, \IEEEmembership{Member, IEEE}

\thanks{This work is performed in the project Master360 under grant 20D1905C, funded by the German Federal Ministry for Economic Affairs and Climate Action within the Luftfahrtforschungsprogramm (LuFo).
}
\thanks{The authors are with the Institute of Microwave and Wireless Systems, Leibniz University Hannover, Appelstr. 9A, 30167 Hannover, Germany \mbox{(e-mail:} \mbox{grundmann@imw.uni-hannover.de;} \mbox{warkentin@imw.uni-hannover.de;} \mbox{manteuffel@imw.uni-hannover.de)}
}}

\maketitle

\begin{abstract}
A method to derive features of modal eigenvalue traces from known and understood solutions is proposed. It utilizes the concept of subduction from point group theory to obtain the symmetry properties of a target structure from those of a structure with a higher order of symmetry. This is applied exemplary to the analytically known characteristic modes (CMs) of the spherical shell. {The eigenvalue behavior of a cube in free-space is derived from it numerically. In this process, formerly crossing eigenvalue traces are found to split up, forming a macroscopic crossing avoidance (MACA). This finding is used to explain indentations in eigenvalue traces observed for 3-D structures, which are of increasing interest in recent literature. The utility of this knowledge is exemplified through a demonstrator antenna design. Here, the subduction procedure is used to analytically predict the eigenvalues of a cuboid on a perfectly electrically conducting plane. The a priori knowledge about the MACA is used to avoid its negative impact on input matching and the frequency stability of the far-field patterns, by choosing the dimensions of the antenna structure so the MACA is outside the target frequency range.} 
\end{abstract}

\begin{IEEEkeywords}
Characteristic modes (CMs), antenna theory, multi port antennas, group theory, symmetry
\end{IEEEkeywords}

%
\IEEEpeerreviewmaketitle

\section{Introduction} \label{sec:Introduction}

\IEEEPARstart{I}{n} recent years, characteristic modes (CMs) \cite{Harrington1971a} have become a vital tool for antenna design. Many works utilize the provided, intuitive insight into the electromagnetic behavior of an antenna structure as a design assistance in communication \cite{Manteuffel2016, Peitzmeier2022, Adams2022, Huang2023, ElYousfi2023} or sensing applications \cite{RanjbarNikkhah2020, Ma2022, Grundmann2023}. Others expand the theoretical and computational background \cite{Schab2016, Schab2017, Masek2020, Gustafsson2022, Capek2023}. The application of group theory, based on the antenna's symmetry properties, recently lead to further advances in both areas. In particular, a design method that utilizes symmetry to reach the upper bound for the number of perfectly uncorrelated ports on an antenna structure is proposed in \cite{Peitzmeier2019, Masek2021, Peitzmeier2022}. Furthermore, a symmetry-based solution to the intensely debated issue of eigenvalue tracking is presented in \cite{Masek2020}. Tracking is required to obtain information on the behavior of a mode across frequency, as the defining eigenvalue problem yields independent solutions for each frequency. In particular, the observation that some modal traces do cross and others do not (crossing avoidance, \cite{Schab2016}), is identified as a consequence of symmetry in  \cite{Schab2017, Masek2020}. {In this paper, the group theory based method of subduction is proposed for use in antenna analysis and design.}

Simultaneously, the variety of applications and antenna structures increases. In addition to \mbox{2-D} setups that represent, for example, patches \cite{Peitzmeier2022, Qin2023, Tian2023}, metasurfaces \cite{Gao2020, ElYousfi2023, Gao2023} or mobile terminals \cite{Hei2021, Li2022, Pang2023}, \mbox{3-D} structures like aircraft \cite{Vogel2017, Wang2019a, Ma2022} and ground vehicles \cite{Ma2018, RanjbarNikkhah2020} are of interest. In the light of the mentioned symmetry-based methods, the recently increased interest in applications with highly symmetric \mbox{3-D} structures is of particular importance. {These include dedicated symmetrical multi mode multi port antennas (M$^3$PAs) \cite{Grundmann2021, Grundmann2023}, as well as platform-based antennas on cubesat \cite{Dicandia2020, Narbudowicz2021, LLangaVargas2022}.} 

\begin{figure}
    \centering
    \subfloat[(a)]{
    \includegraphics[scale=0.48]{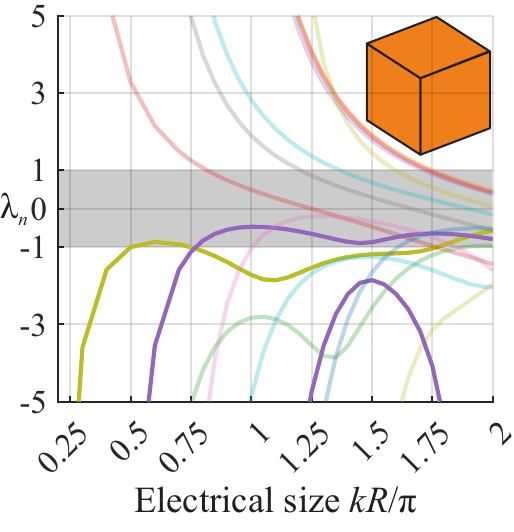}
    \label{fig:MotivationEVCube}}
    \hfil
    \subfloat[(b)]{
    \includegraphics[scale=0.48]{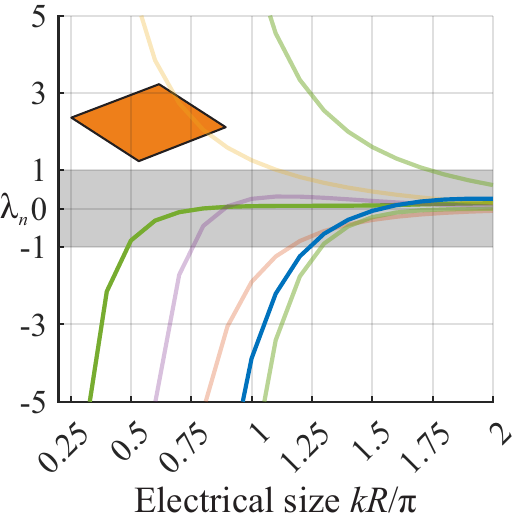}
    \label{fig:MotivationEVPlate}} 
    \caption{CM eigenvalues $\lambda_n$ of selected modes of a (a) cube and a (b) square patch in free-space over electrical size $kR/\pi$. For an electrical size of $kR/\pi=1$, the diameter of the smallest sphere circumscribing the structure is one wavelength. }
    \label{fig:Motivation}
\end{figure}
{In Fig.~\subref*{fig:MotivationEVCube}, the eigenvalue traces for a set of CMs on such a \mbox{3-D} structure are depicted. The CM eigenvalue traces of the cube are relatively complex. It is observed that the curvature of some traces changes between positive and negative. As an example, refer to the highlighted olive and purple colored lines in Fig.~\subref*{fig:MotivationEVCube}. Visually, this effect appears as an indentation of varying depth in the eigenvalue trace. The effect is typical for many \mbox{3-D} structures and observable in various works from the literature \cite{Chen2018, Masek2020, LLangaVargas2022}.
In contrast, the eigenvalue traces of a \mbox{2-D} square plate, shown in Fig.~\subref*{fig:MotivationEVPlate}, all have either positive or negative curvature. This simple and predictable behavior is considered typical for various \mbox{2-D} structures and is observed in many works \cite{Manteuffel2016, Safin2016, Adams2022}. 

These observations have consequences for antenna design: In order to achieve broadband input matching for a port assigned to the $n$th CM, that CM is commonly expected to have an eigenvalue $\lambda_n$ close to zero across the desired frequency range \cite{CabedoFabres2007}. If the indentations observed for \mbox{3-D} structures are not addressed in an antenna design process, they can lead to a deteriorated input matching in sections of the target frequency band. 
Consequently, an intuitive interpretation of this effect is desirable, that allows the prediction of its occurrence and consequences in the early stages of an antenna design process. Measures to reduce its impact are to be developed on this basis.

To this end, this paper introduces the concept of subduced representations to antennas, based on the recent advances in the application of group theory. The method is presented in Section~\ref{sec:Sphere}, along with some necessary background. In Section~\ref{sec:changeSymmetryGroup}, it is employed to explain the complex eigenvalue behavior of \mbox{3-D} structures, using the analytical solutions available for the spherical shell. 
Section~\ref{sec:geometryModifications} explores how eigenvalue traces behave if the sphere is modified towards a cube, providing the desired intuitive explanation for the behavior observed in Fig.~\subref*{fig:MotivationEVCube}. 

We used the proposed procedure to assist in the design of the demonstrator antenna first presented in \cite{Grundmann2023}, but did not include it in that publication, since it focused on the evaluation of direction-finding antennas. Therefore, the design aspects of the demonstrator antenna related to the novel subduction procedure are first detailed in Section~\ref{sec:antennaDevelopment}. 
For completeness, the absence of indentations in the eigenvalue traces of the \mbox{2-D} square plate from Fig.~\subref*{fig:MotivationEVPlate} is explained in Appendix~\ref{sec:cubeToPlanar}.}

\section{Method to Derive CM Eigenvalue Behavior From Established Solutions} \label{sec:Sphere}
The most common formulation used to determine CMs is a generalized eigenvalue problem expressed in terms of method of moments (MoM) quantities, expanded by a set of basis functions \cite{Harrington1971a} 
\begin{equation} \label{eq:GEP}
    \mathbf{XI}_n = \lambda_n \mathbf{RI}_n ~.
\end{equation}
Here, $\mathbf{X}$ and $\mathbf{R}$ are the imaginary and real parts of the MoM impedance matrix $\mathbf{Z}$, respectively. The vectors $\mathbf{I}_n$ are typically referred to as the CM eigencurrents, while $\lambda_n$ are the corresponding eigenvalues. Numerical results throughout this article are determined for each frequency using the Leibniz University Hannover (LUH) in-house MATLAB implementation of (\ref{eq:GEP}), which is evaluated in \cite{Chen2018}. 

\subsection{Established Solution: CMs of the Spherical Shell} \label{sec:analyticalSolution}
For some geometries, an analytic representation of the CMs is known. 
In this work, we utilize the complex, but analytically known and understood CMs of a perfectly electrically conducting (PEC) spherical shell \cite{Garbacz1968}.  
{These are used as a reference to derive the eigenvalue behavior of structures with lower symmetry, using an analytical approach in Section~\ref{sec:changeSymmetryGroup}, assisted by numerical simulations in Section~\ref{sec:geometryModifications}.}
However, the proposed derivation method is not limited to a specific analytical solution. Any well-understood CM solution can be used instead, as is demonstrated in Section~\ref{sec:antennaDevelopment}.

For the spherical shell, the CMs are identical to the spherical wave functions (SWFs), also known as spherical harmonics. 
Their eigenvalues are given by \cite{Garbacz1968} 
\begin{equation} \label{eq:sphereEigenvalues}
    \lambda_{t,s} = 
    \begin{cases} - \frac{y_t(kR)}{j_t(kR)} , & \mathrm{for}~ s=1 ~\mathrm{(TE)} \\
    - \frac{\frac{\partial}{\partial(kr)} (kry_t(kr)) \vert_{r=R}} {\frac{\partial}{\partial(kr)} (krj_t(kr)) \vert_{r=R}} , & \mathrm{for}~ s=2 ~\mathrm{(TM).}
    \end{cases} 
\end{equation}
Here, $k$ is the free-space wavenumber, and $R$ the radius of the sphere. $j_t(kR)$ and $y_t(kR)$ are the spherical Bessel functions of order $t$ of the first and second kinds (Neumann function), respectively. Modes with $s=1$ are of the transverse electric (TE) type. Those with $s=2$ are of the transverse magnetic (TM) type. 

\begin{figure}
    \centering
    \includegraphics[scale=1]{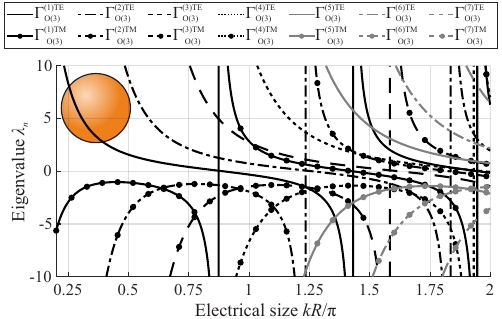}
    \caption{Eigenvalues of CMs on the spherical PEC shell with $t\leq7$ according to (\ref{eq:sphereEigenvalues}). The CMs of each trace are basis functions of a different irrep $\Gamma^{(t),s}_\mathrm{O(3)}$ of the $O(3)$ symmetry group (legend).}
    \label{fig:EVSphere}
\end{figure}
The CM eigenvalue traces for the spherical shell are determined from (\ref{eq:sphereEigenvalues}) and shown in Fig.~\ref{fig:EVSphere} for $kR/\pi\leq2$. This corresponds to a diameter of two free-space wavelengths, which covers most applications of CM antenna design. Apparently, the electrical size $kR/\pi$ is proportional to both the frequency and the radius of the sphere. All three terms are used interchangeably in this work for describing qualitative developments. 

The global index $n$ in $\lambda_n$ is determined from $t$ {and $s$} using the formulation given in \cite{Hansen1988} 
\begin{equation} \label{eq:multiindex}
    n = 2\left( t(t+1) + m - 1 \right) + s ~,~ -t\leq m\leq t ~.
\end{equation}
The eigencurrents and far-fields of the modes are given in \cite{Garbacz1968}. The presence of the integer $m$ indicates that multiple CMs share the same eigenvalue for all frequencies. These degeneracies appear in different orders, which are given by \cite{Cornwell1997} 
\begin{equation} \label{eq:degreeOfDegeneracy}
    M = 2t+1 ~, 
\end{equation}
depending on $t$. {This means that the modes represented by the black solid curves with (TM) and without (TE) marker in Fig.~\ref{fig:EVSphere} are threefold degenerate ($t=1$). For $t=2$ (black dashed-dotted lines in Fig.~\ref{fig:EVSphere}), the modes are fivefold degenerate and so on.} In order to understand the impact of this observation, a brief recap of symmetry and group theory with respect to CMs is required. 

\subsection{Tools: Symmetry and the Von Neumann-Wigner Theorem} \label{sec:symmetry}
The symmetry group of a structure is a set of symmetry operations $T$ under which the structure is invariant \cite{Peitzmeier2022}. These symmetry operations, like rotations, can be used to transform currents on the structure. If this set of operations is expressed as a set of matrices, it is called a matrix representation ${\Gamma}$ of the group \cite{Masek2020}. Since there is an infinite number of bases in which a current can be expressed, e.\,g., CMs, there is also an infinite number of matrix representations. However, representations can be reduced to a set of irreducible representations (irreps), which is unique for each group up to a similarity transformation. 
The trace of the irrep matrix $\mathbf{\Gamma}(T)$ for a symmetry operation $T$ is called the character $\chi(T)$.
Now, any current $\mathbf{I}$ on the structure can be projected onto an irrep {$\Gamma^{(p)}$}, to determine how much of it belongs to this irrep, using \cite{Cornwell1997} 
\begin{equation} \label{eq:projectCurrentOnIrrep}
    \mathcal{P}^{(p)} \mathbf{I} = w^{(p)} \boldsymbol\psi^{(p)} ~.
\end{equation}
Here, $\mathcal{P}^{(p)}$ is the character projection operator \cite{Cornwell1997} and $w^{(p)}$ a weighting coefficient.
If a current only belongs to a single irrep {$\Gamma^{(p)}$ with $w^{(p)}=\delta_p$}, it is called a basis function $\boldsymbol\psi^{(p)}$ of {$\Gamma^{(p)}$. Here, $\delta_p$ is the Dirac delta function}. In this case, all projections to other irreps yield zero. 
For each irrep, $d_p$ mutually orthogonal basis functions can be found, where $d_p$ is the dimension of the irrep. 
CMs are basis functions of the irreps of the symmetry group of the investigated structure \cite{Peitzmeier2019}. As such, CMs are uncorrelated to other CMs if they belong to a different irrep, or if they are mutually orthogonal basis functions from the same, multidimensional irrep. 
{Since this also holds for port currents, each irrep can be assigned exactly $d_p$ antenna ports that are orthogonal to each other and to all those assigned to other irreps.}
{This observation is utilized to generate uncorrelated antenna ports in Section~\ref{sec:antennaDevelopment} \cite{Peitzmeier2022, Grundmann2023}.}
{If basis functions are known, the entries of the irrep matrix $\mathbf{\Gamma}^{(p)}(T)$ are determined by \cite{Cornwell1997} 
\begin{equation} \label{eq:irrepDefinition}
    \Gamma_{\mu,\nu}^{(p)}(T) = \langle \boldsymbol\psi^{(p)}_\mu, P(T)\boldsymbol\psi^{(p)}_\nu \rangle 
\end{equation}
where $\langle\cdot,\cdot\rangle$ is the inner product and $P(T)$ is the transformation operator associated with the symmetry operation $T$, while $\mu$ and $\nu$ are the row and column indices, respectively.}

The von Neumann-Wigner theorem
states that the dimension of the irrep equals the order of degeneracy of a mode belonging to that irrep \cite{Neumann1929,Masek2020}.
Importantly, the von Neumann-Wigner theorem also holds for accidental degeneracies. Accidental degeneracies appear only at discrete frequencies, not across the complete frequency range. Since the crossing of two eigenvalue traces always involves an accidental degeneracy, a crossing of eigenvalue traces from the same irrep would result in a higher degeneracy and is therefore not possible \cite{Masek2020}. This behavior is known in the literature as a crossing avoidance \cite{Schab2016, Schab2016a}. 
Recently, attempts were made to address antenna design implications of crossing avoidance through geometry modifications \cite{Elias2022}, promoting the need for a systematic approach in this regard. 

Applying these findings to the spherical shell, we find that its symmetry group is the orthogonal group in three dimensions $O(3)$. 
In contrast to most practically relevant groups for antenna design, $O(3)$ is an infinite group. It consists of an infinite number of symmetry operations and has an infinite number of irreps. 
The dimensions of the irreps increase with $t$ and are given by (\ref{eq:degreeOfDegeneracy}) \cite{Cornwell1997}. 
For every $t$, there is an irrep for TE modes ${\Gamma}^{(t)\mathrm{TE}}_\mathrm{O(3)}$, and one for TM modes ${\Gamma}^{(t)\mathrm{TM}}_\mathrm{O(3)}$.
Consequently, exactly $d_p = M$ degenerate modes belong to any irrep. Accordingly, all modes from an irrep always have the same eigenvalue and there are no limitations to the crossing of eigenvalue traces for the $O(3)$ group in Fig.~\ref{fig:EVSphere}. 

Of particular importance for the understanding of modal eigenvalue behavior is the presence of poles in the eigenvalue traces of all modes on the spherical shell. In Fig.~\ref{fig:EVSphere}, these are indicated by vertical lines. It is important to note that these vertical lines do not indicate a crossing of traces since the eigenvalue at the frequency of the pole is negative infinity, if it is approached from a lower frequency, or positive infinity, if it is approached from a higher frequency. However, almost all other traces are crossed in the vicinity of the pole, during the descend
to negative infinity and the descend from positive infinity. {Refer to the black solid trace with circular marker in Fig.~\ref{fig:EVSphere} as an example. 
Next, we utilize this background knowledge to link this analytical CM solution to CMs of a structure with lower symmetry.}

\subsection{Method: Derivation of Other Structures by Modifying the Symmetry Group Using Subduction} \label{sec:changeSymmetryGroup}
As indicated above, practically relevant antenna structures are of finite symmetry. This is evident since any port setup must be of finite symmetry \cite{Grundmann2022a}. In fact, it is not possible to determine the exact eigenvalues of a structure with infinite symmetry using numerical methods that rely on discretization, like MoM. Any mesh applied to the spherical shell can only achieve finite symmetry, due to the finite number of mesh cells. {Therefore, by applying a mesh to the spherical shell, the $O(3)$ symmetry group dissolves into one of its subgroups. In order to establish and understand the connection between the numerical results obtained for meshes of finite symmetry and the analytical CM solution of the spherical shell, a method is required to predict the consequences of a change of the symmetry group. 

This method is called subduction, as the representations of a subgroup are said to be subduced by the representations of the original group \cite{Ludwig1996}.} 
Subduction is a well-established concept in chemistry \cite{Ludwig1996}, 
but, to the authors' best knowledge, has not been applied yet to antennas. 
When a group $\mathcal{H}$ is dissolved into one of its subgroups $\mathcal{G}$, its irreducible representations become reducible representations $\mathbf{\Gamma}_\mathcal{G}(T)$ of $\mathcal{G}$. 
The number of times an irreducible representation $\mathbf{\Gamma}^{(p)}_\mathcal{G}(T)$ is contained in the reducible representation $\mathbf{\Gamma}_\mathcal{G}(T)$ is determined by \cite{Cornwell1997} 
\begin{equation} \label{eq:subduction}
    n_p = \frac{1}{g}\sum\limits_{T\in\mathcal{G}} \chi(T)  \chi^{(p)}(T)^* ~.
\end{equation}
Here, $g$ is the number of symmetry operations $T$ in the group and $\chi(T)$ and $\chi^{(p)}(T)$ are the characters of $\mathbf{\Gamma}_\mathcal{G}(T)$ and $\mathbf{\Gamma}^{(p)}_\mathcal{G}(T)$, respectively. 

\begin{figure}
    \centering
    \subfloat[(a)]{
    \includegraphics[]{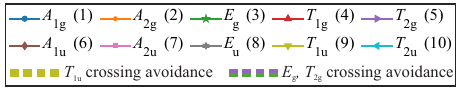}
    \label{fig:accuracySph:legend}}
    \hfil
    \subfloat[(b)]{
    \includegraphics[width=0.45\columnwidth]{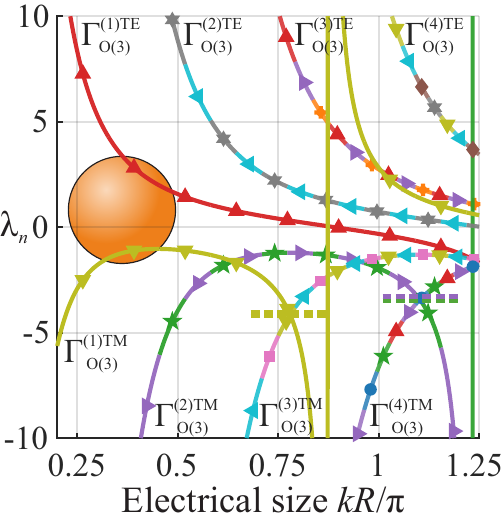}
    \label{fig:accuracySph:analytical}}
    \hfil
    \subfloat[(c)]{
    \includegraphics[width=0.45\columnwidth]{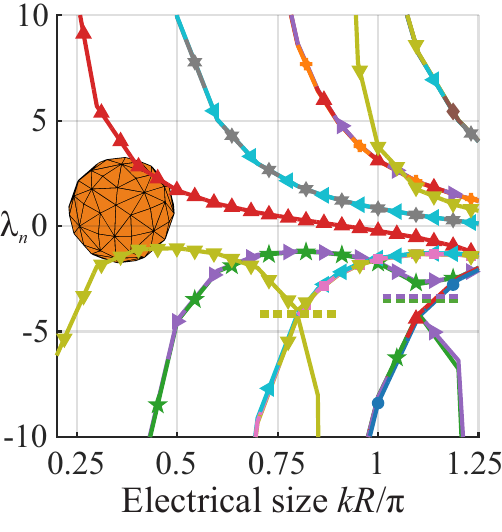}
    \label{fig:accuracySph:numerical}}
    \caption{Eigenvalue traces of the spherical PEC shell, assigned to $O_\mathrm{h}$ irreps, demonstrating the accuracy of the proposed subduction method. \protect\subref{fig:accuracySph:analytical} Analytical eigenvalues, irrep assignment and crossing avoidance prediction using subduction (\ref{eq:subduction}). \protect\subref{fig:accuracySph:numerical} Numerical MoM eigenvalues, irrep assignment and crossing avoidance determination using the conventional method (\ref{eq:projectCurrentOnIrrep}). \protect\subref{fig:accuracySph:legend}: Legend.}
    \label{fig:accuracySph}
\end{figure}

{To illustrate this procedure, we determine the irreps of the octahedral group $\mathcal{G}=O_\mathrm{h}$, as an exemplary subgroup of $\mathcal{H}=O(3)$, through subduction. 
Throughout this work, the Schoenflies notation \cite{Cornwell1997} is used to name finite groups. 
The $O_\mathrm{h}$ group is practically relevant since it is the symmetry group of the cube, which is a common structure for antenna systems like cubesat. However, the utilized method of investigation is easily applicable to other point groups, like the icosahedral group $I_\mathrm{h}$. For now, it is important to keep in mind that the investigated structure is still an approximate sphere.

The $O_\mathrm{h}$ group has ten irreps, which are given in Fig.~\subref*{fig:accuracySph:legend}.
In addition to the names, the irrep numbers $p$ are given in parenthesis in Fig.~\subref*{fig:accuracySph:legend}. Both are taken from \cite{Cornwell1997}, where corresponding data are also provided.
The $A_\mathrm{\lbrace 1,2 \rbrace\lbrace u,g \rbrace}$ irreps are 1-D, the $E_\mathrm{\lbrace u,g \rbrace}$ irreps are 2-D and the $T_\mathrm{\lbrace 1,2 \rbrace\lbrace u,g \rbrace}$ irreps are 3-D.} 

As stated above, each trace in Fig.~\ref{fig:EVSphere} belongs to a unique irrep in $O(3)$. Therefore, the $O_\mathrm{h}$ irreps of a trace are subduced from the respective $O(3)$ irrep. These relations are determined using (\ref{eq:subduction}) and given in Table~\ref{tab:OhSubducedByO3}. If an irrep is contained more than once, this is indicated by the number $n_p$ in parenthesis. For example, ${T_\mathrm{1g}}(2)$ indicates that the ${T_\mathrm{1g}}$ irrep is contained $n_p=2$ times. 
{The subduced $O_\mathrm{h}$ irreps are assigned to the analytically known eigenvalue traces of the spherical shell in Fig.~\subref*{fig:accuracySph:analytical}. This display is obtained using only (\ref{eq:sphereEigenvalues}), (\ref{eq:subduction}) and the characters of the two groups. It provides a computationally efficient and accurate prediction of the numerical results, which are depicted in Fig.~\subref*{fig:accuracySph:numerical}. Here, eigenvalues were obtained using a MoM simulation (\ref{eq:GEP}) and assigned to the irreps using the conventional projection operator method (\ref{eq:projectCurrentOnIrrep}).} 

From {Fig.~\ref{fig:accuracySph}} and Table~\ref{tab:OhSubducedByO3}, it is observed that the modes of the first TE irrep ${\Gamma}^{(1)\mathrm{TE}}_\mathrm{O(3)}$ are uniquely mapped to ${T_\mathrm{1g}}$, while the modes of ${\Gamma}^{(1)\mathrm{TM}}_\mathrm{O(3)}$ are assigned in the same fashion to ${T_\mathrm{1u}}$. This unique mapping is possible since both irreps have the same dimension, respectively. For traces from any other irreps of $O(3)$ however, a mapping to a unique irrep of $O_\mathrm{h}$ is not possible. This is obvious since their dimensions exceed the maximum dimension of the $O_\mathrm{h}$ irreps, see (\ref{eq:degreeOfDegeneracy}). 
Consequently, these traces have multiple colors in {Fig.~\ref{fig:accuracySph}}. For example, the five degenerate modes from the second TM irrep of $O(3)$, ${\Gamma}^\mathrm{(2)TM}_\mathrm{O(3)}$, are mapped onto ${E_g}$ and ${T_\mathrm{2g}}$ (purple-green trace). 

{Through the accurate assignment of irreps, the subduction method also predicts the occurence of crossing avoidances.}
{Following the von Neumann-Wigner theorem, traces belonging to the same irrep of the new symmetry group $O_\mathrm{h}$ are no longer allowed to cross.
As an example, take the modes of ${\Gamma}^\mathrm{(3)TM}_\mathrm{O(3)}$ (olive-cyan-pink trace), whose eigenvalues cross those of ${\Gamma}^\mathrm{(1)TM}_\mathrm{O(3)}$ (olive trace) near $kR/\pi=0.75$ in Fig.~\subref*{fig:accuracySph:analytical}. The crossing avoidances are highlighted here using dashed lines.} 
Since both traces contain modes belonging to ${T_\mathrm{1u}}$, the crossing is not allowed for these modes under $O_\mathrm{h}$ symmetry. Consequently, the modes swap the position of their eigenvalue traces and their modal currents after this crossing avoidance. {In particular, an indentation and a peak are created in the two traces, respectively. By investigating Fig.~\subref*{fig:accuracySph:numerical}, we find that the numerically determined locations of these crossing avoidances and the modes involved are indeed accurately predicted by Fig.~\subref*{fig:accuracySph:analytical}. Differences in the eigenvalues are due to the limited resolutions of frequency and mesh in the numerical simulation.}

{Since the geometry of the investigated structure is not modified in this example, the appearance of crossing avoidances is only a numerical effect with no practical consequences for the electromagnetic behavior of the structure. However, the subduction of irreps constitutes the fundamental principle that determines the behavior of CM eigenvalues and far-fields of various other geometries, as will become evident in Sections~\ref{sec:geometryModifications}, \ref{sec:antennaDevelopment} and the Appendixes.} 

\begin{table}
\renewcommand{\arraystretch}{1.6}
\caption{$O_\mathrm{h}$ Irreps Subduced by $O(3)$ Irreps: \\How Often Are $O_\mathrm{h}$ Irreps Contained in $O(3)$ Irreps}
\label{tab:OhSubducedByO3}
\centering
\begin{tabular}{|c|c|c|}
    \hline
    $O(3)$ irrep & $d_p=M$ & $O_\mathrm{h}$ irreps ($n_p$)\\ 
    \hline\hline
${\Gamma}^{(1)\mathrm{TE}}_\mathrm{O(3)}$ & 3 & ${T_\mathrm{1g}}$\\
\hline
${\Gamma}^{(1)\mathrm{TM}}_\mathrm{O(3)}$ & 3 & ${T_\mathrm{1u}}$\\
\hline
${\Gamma}^{(2)\mathrm{TE}}_\mathrm{O(3)}$ & 5 & ${E_\mathrm{u}},{T_\mathrm{2u}}$\\
\hline
${\Gamma}^{(2)\mathrm{TM}}_\mathrm{O(3)}$ & 5 & ${E_\mathrm{g}},{T_\mathrm{2g}}$\\
\hline
${\Gamma}^{(3)\mathrm{TE}}_\mathrm{O(3)}$ & 7 & ${A_\mathrm{2g}}, {T_\mathrm{1g}}, {T_\mathrm{2g}}$\\
\hline
${\Gamma}^{(3)\mathrm{TM}}_\mathrm{O(3)}$ & 7 & ${A_\mathrm{2u}}, {T_\mathrm{1u}}, {T_\mathrm{2u}}$\\
\hline
${\Gamma}^{(4)\mathrm{TE}}_\mathrm{O(3)}$ & 9 & ${A_\mathrm{1u}}, {E_\mathrm{u}}, {T_\mathrm{1u}}, {T_\mathrm{2u}}$\\
\hline
${\Gamma}^{(4)\mathrm{TM}}_\mathrm{O(3)}$ & 9 & ${A_\mathrm{1g}}, {E_\mathrm{g}}, {T_\mathrm{1g}}, {T_\mathrm{2g}}$\\
\hline
${\Gamma}^{(5)\mathrm{TE}}_\mathrm{O(3)}$ & 11 & ${E_\mathrm{g}}, {T_\mathrm{1g}} (2), {T_\mathrm{2g}}$\\
\hline
${\Gamma}^{(5)\mathrm{TM}}_\mathrm{O(3)}$ & 11 & ${E_\mathrm{u}}, {T_\mathrm{1u}} (2), {T_\mathrm{2u}}$\\
\hline
${\Gamma}^{(6)\mathrm{TE}}_\mathrm{O(3)}$ & 13 & ${A_\mathrm{1u}}, {A_\mathrm{2u}}, {E_\mathrm{u}}, {T_\mathrm{1u}}, {T_\mathrm{2u}} (2)$\\
\hline
${\Gamma}^{(6)\mathrm{TM}}_\mathrm{O(3)}$ & 13 & ${A_\mathrm{1g}}, {A_\mathrm{2g}}, {E_\mathrm{g}}, {T_\mathrm{1g}}, {T_\mathrm{2g}} (2)$\\
\hline
\vdots & \vdots & \vdots\\
\hline
\end{tabular}
\end{table}

{
\section{Impact of Geometry Modifications on Subduced Irreps} \label{sec:geometryModifications}
In the following, it is established that crossing avoidances caused by the subduction of irreps evolve from a numerical artifact to a practical feature with impact on antenna design, if the geometry is modified. 
This finding will affect both input matching and far-field characteristics of the antenna ports eventually designed in Section~\ref{sec:implementation}.
The second objective is to establish a direct mapping of the modes on the sphere to those on the cube. 
The design process of a practical antenna subduced from the cube would thereby benefit from the link to the analytical solutions for the CMs on the spherical shell, as we show in Section~\ref{sec:antennaDevelopment}. }

\begin{figure}
    \centering
    \subfloat[(a)]{
    \includegraphics[scale=0.5]{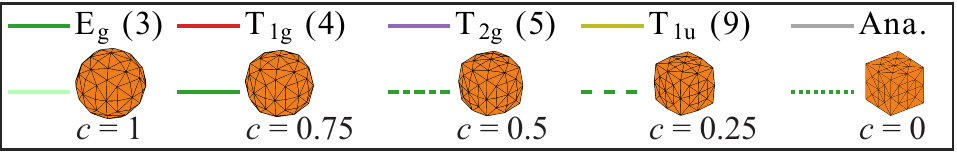}
    \label{fig:sph2cube:legend}}
    \hfil
    \subfloat[(b)]{
    \includegraphics[width=0.45\columnwidth]{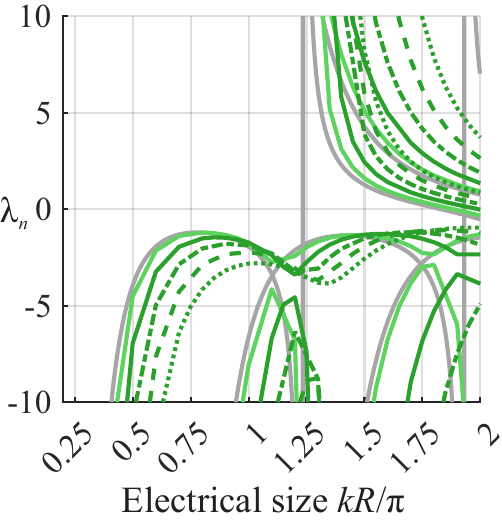}
    \label{fig:sph2cube:irrep3}}
    \hfil
    \subfloat[(c)]{
    \includegraphics[width=0.45\columnwidth]{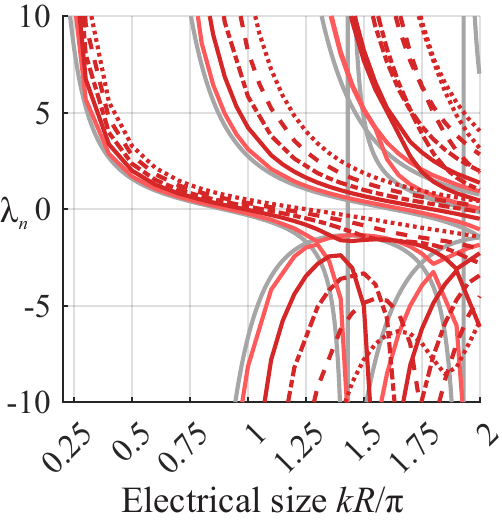}
    \label{fig:sph2cube:irrep4}}
    \hfil
    \subfloat[(d)]{
    \includegraphics[width=0.45\columnwidth]{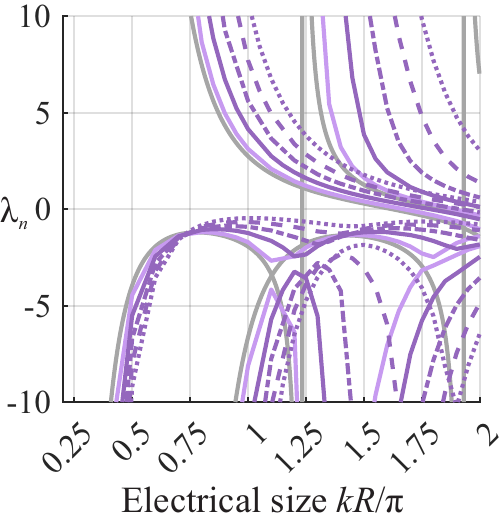}
    \label{fig:sph2cube:irrep5}}
    \hfil
    \subfloat[(e)]{
    \includegraphics[width=0.45\columnwidth]{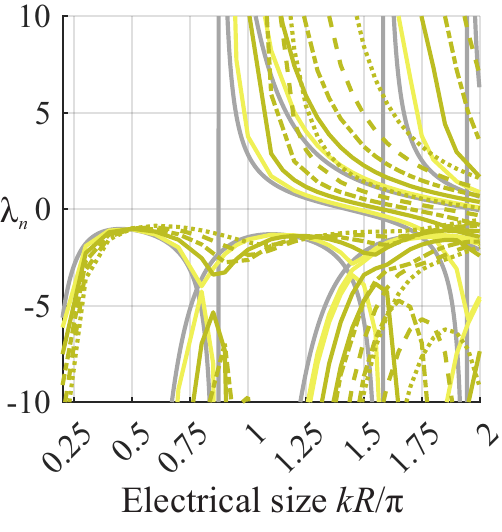}
    \label{fig:sph2cube:irrep9}}
    \caption{Derivation of MACAs using eigenvalue traces for different sphericities (parameter $c$) of the antenna, see \protect\subref{fig:sph2cube:legend}. Analytical results for the sphere are depicted in gray, the geometries are identified by line style and brightness. \protect\subref{fig:sph2cube:irrep3} - \protect\subref{fig:sph2cube:irrep9} contain selected CM eigenvalue traces belonging to the $O_\mathrm{h}$ irreps ${E_\mathrm{g}}$, ${T_\mathrm{1g}}$, ${T_\mathrm{2g}}$ and ${T_\mathrm{1u}}$, respectively.}
    \label{fig:sph2cube}
\end{figure}

The simulation model of the cube is depicted on the right-hand side of Fig.~\subref*{fig:sph2cube:legend}. Each face is divided along its symmetry axes into eight congruent triangles. Each of these triangles is subdivided into three triangles. 
Now, the cube is inflated by modifying the distance $b_c$ of each vertex of this triangular grid to the center of the cube. This is done by modifying the parameter $0\leq c\leq 1$ in
\begin{equation}
    b_c = (R-b_0)c+b_0 ~.
\end{equation}
Here, the original distance between the respective vertex on the cube and the center is given by $b_0$, and the radius of the smallest sphere circumscribing the cube is $R$. Depictions of the resulting geometries are given in Fig.~\subref*{fig:sph2cube:legend}. 

Each of the five resulting structures is simulated numerically with varying electrical size. 
To connect the individual eigenvalues to traces in Fig.~\ref{fig:sph2cube}, a correlation-based tracking algorithm is employed, and adherence to the von Neumann-Wigner theorem is enforced as postprocessing.%
\footnote{Sometimes, this procedure fails since traces leave the investigated eigenvalue range \cite{Masek2020}. Fortunately, an analytic ground truth is available for the spherical shell and the geometry is only modified gradually. Therefore, manual correction is straightforward.}
The results are depicted in Fig.~\ref{fig:sph2cube}, together with the respective analytical results for the spherical shell (gray lines). To improve the clarity of the visualization, some higher order modes are omitted.

In the following, only the results for the irreps ${E_\mathrm{g}}$, ${T_\mathrm{1g}}$, ${T_\mathrm{2g}}$, and ${T_\mathrm{1u}}$ are investigated. This is possible since modes from different irreps are independent of each other. As can be seen from Fig.~\ref{fig:sph2cube}, investigating irreps independently provides a better overview than, for example, Fig.~\ref{fig:EVSphere}.
{The selected irreps are chosen since crossing avoidances, along with indentations and peaks in eigenvalue traces, are observable here for small electrical sizes.} 
Small electrical sizes are particularly relevant since, here, a smaller number of modes is near significance ($|\lambda_n|\leq 1$). If an indentation in a modal trace is observed for greater electrical sizes, commonly a different mode from the same irrep remains significant, reducing the adverse impact on port input matching. For an example, take the {dotted traces in Fig.~\subref*{fig:sph2cube:irrep5}.}

{As discussed for Fig.~\ref{fig:accuracySph}, the eigenvalues obtained for $c=1$ are still very close to those of the spherical shell. 
The crossing avoidance of the first- and second-order TM modes is visible near $kR/\pi=1.1$ in Fig.~\subref*{fig:sph2cube:irrep3} and \subref*{fig:sph2cube:irrep5} (brighter lines). For simplicity, we call the effect of the crossing avoidance on the first mode an eigenvalue indentation and its effect on the second mode an eigenvalue peak. 
With decreasing sphericity and decreasing $c$, the depth of the indentation also decreases.} 
This is a general trend for all irreps. In Fig.~\subref*{fig:sph2cube:irrep3}, this effect overlaps with a general decrease of the eigenvalues of the first TM mode.
The decreasing depth becomes evident if the indentations are interpreted as remnants of the poles found in the eigenvalue traces of the spherical shell. In this sense, altering this structure influences the impact of these poles on the CM eigenvalue behavior.
{Interestingly, the relative position of the indentation and the peak is maintained, and they remain aligned to each other.} This underlines that both traces originate from the same crossing avoidance. 

Furthermore, the eigenvalue peaks move toward higher eigenvalue magnitudes as the sphericity decreases in Fig.~\subref*{fig:sph2cube:irrep3}, \subref*{fig:sph2cube:irrep4} and \subref*{fig:sph2cube:irrep9}. Note that the exact peak values are altered by the discretization of the sweep over $kR/\pi$, but the general trend is evident. 
{Consequently, the distance between the eigenvalues at the peak and those at the indentation increases as $c$ decreases. We call this observation a macroscopic crossing avoidance (MACA), as it affects macroscopic parameters like port input matching. In this sense, the type of crossing avoidance observed for $c=1$ is a microscopic crossing avoidance (MICA), which is caused by microscopic changes in the geometry or mesh%
\footnote{The term \textit{numerically induced} is used in \cite{Schab2017} and \cite{Schab2016a} to describe crossing avoidances that are transformed into crossings by applying a symmetrical mesh. Since this definition does not include geometry modifications, a different term is required for the design-driven approach of this paper. }%
.} 
Typical examples for crossing avoidances found in the literature, for example \cite{Schab2016, Masek2020}, are MICAs. 
If significant modifications of the geometry are conducted, the MICA induced by the von Neumann-Wigner theorem transforms into an MACA. In particular, the von Neumann-Wigner theorem thereby explains and predicts the behavior of the modified geometry based on an analytical solution for the original geometry, using the {subduction} method proposed in Section~\ref{sec:Sphere}. 

To highlight how a crossing avoidance on the sphere governs the course of the eigenvalue traces of the cube, consider the following thought experiment. 
If the crossing of the modes in e.\,g. Fig.~\subref*{fig:sph2cube:irrep9} was allowed, one could expect that the crossing point remains at the same eigenvalue for both the sphere and the cube. Instead, the two traces actually split up and the first mode moves closer toward $\lambda_n=0$. 
\begin{figure}
    \centering
    \includegraphics[scale=0.5]{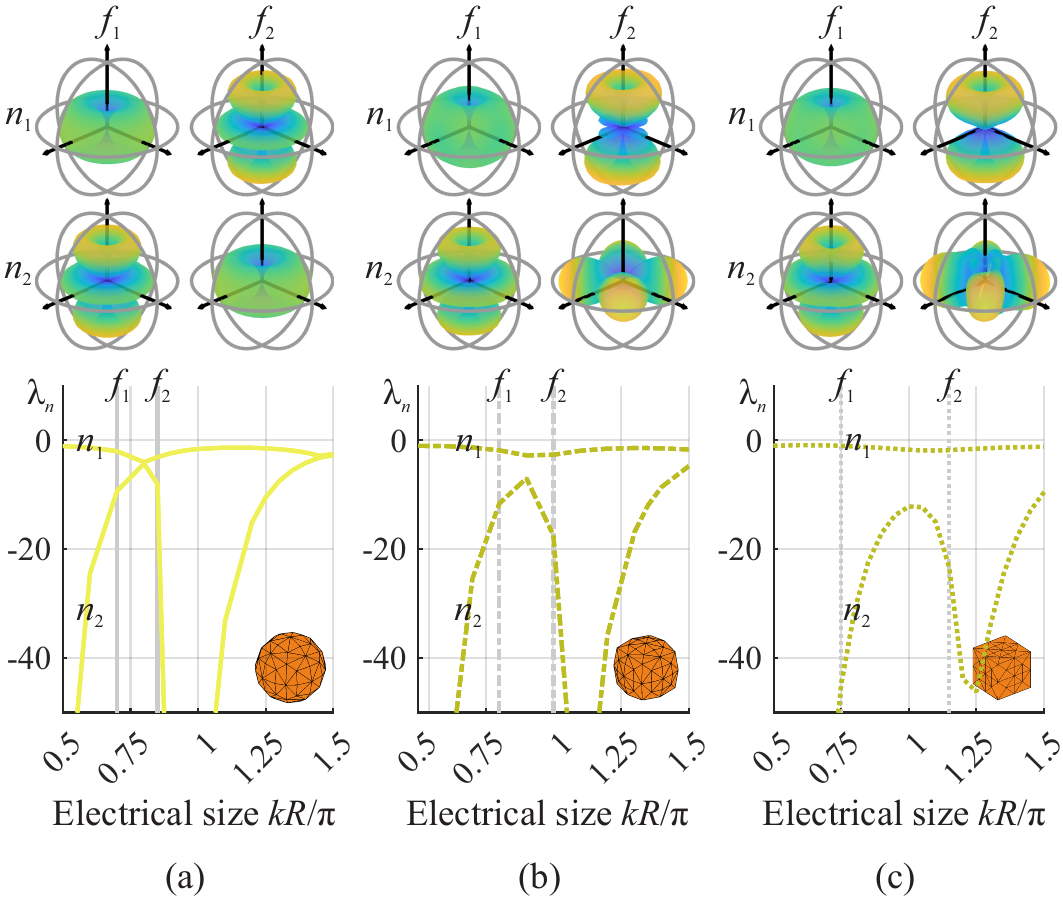}
    \caption{Eigenvalues for the first and second CM traces, $n_1$ and $n_2$, of the irrep ${T_\mathrm{1u}}$. The parameter $c$ varies between (a) - (c), compare Fig.~\ref{fig:sph2cube:irrep9}. Far-field directivities before ($f_1$) and after ($f_2$) the crossing avoidance show that modes still influence each other as the MICA transforms to a MACA and the difference in eigenvalue becomes larger. 
    Only one of the threefold degenerate modes is shown, respectively. A colorbar is given in Fig.~\ref{fig:EVantenna}. }
    \label{fig:farFieldTransition}
\end{figure}
Through the MACA, the two traces still influence each other. This can be considered \textit{a priori} knowledge and is particularly helpful to explain and predict a change in CM far-fields, based only on the eigenvalue traces. Since we established above that indentations and peaks maintain their relative position, the effect can also be identified if only the indentation is observable in the depicted eigenvalue range, like the olive trace in Fig.~\subref*{fig:MotivationEVCube}. 

An example of the change of far-fields due to MACA is depicted in Fig.~\ref{fig:farFieldTransition}(c). It is observed that far-fields are partly interchanged between traces while moving across the crossing avoidance. This effect is well-known for the established MICA \cite{Schab2016}, see Fig.~\ref{fig:farFieldTransition}(a). Additionally, this is superimposed with a general modification of the patterns over frequency. This is attributed to the larger bandwidth across which the crossing avoidance occurs, in comparison to most MICAs. For the second trace $n_2$, the following crossing avoidance is also expected to influence the far-field pattern. 
{Some additional details on the applicability of these observations to different structures are provided in Appendixes~\ref{sec:wireStructures} and \ref{sec:otherStructures}.}


\section{Application to Antenna System Development} \label{sec:antennaDevelopment}
\begin{figure}
    \centering
    \includegraphics[scale=1]{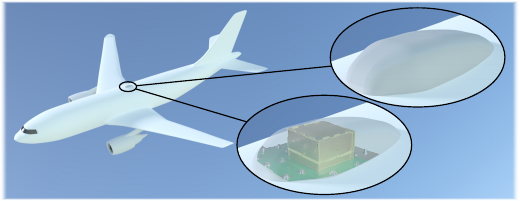}
    \caption{Illustration of the implementation scenario of the proposed antenna, first presented in \cite{Grundmann2023}. Antenna and radome are to be mounted to the aircraft fuselage.}
    \label{fig:plane}
\end{figure}

In {the following, the utility of the proposed subduction procedure and the a priori knowledge on MACAs is demonstrated through an exemplary real-world antenna design. Note that this does not limit the general validity of the proposed subduction procedure to a specific example. 
The demonstrator antenna shall be installed on an aircraft to detect signals transmitted by other aircraft and estimate their directions of arrival. In particular, messages from the airborne collision avoidance system (ACAS) in a frequency range between 1030\,MHz and 1090\,MHz are of interest. The antenna is to be mounted to the aircraft fuselage underneath a radome, as depicted in Fig.~\ref{fig:plane}. For simplicity, the electrically large aircraft body is represented by an infinite PEC plane, and a cuboid is chosen as the fundamental structure of the antenna. This aims for a simpler manufacturing process in comparison to structures conformal to the radome. A patch configuration is not suited since the polarization is required to be perpendicular to the PEC plane. In the following, we determine whether a square or a rectangle is more suited as the footprint of the cuboid antenna, compare Fig.~\ref{fig:cuboidSubduction}(d) and \ref{fig:cuboidSubduction}(e).} 

{\subsection{Subduction of Cuboid on PEC Plane}
\label{sec:subductionToC4v}
Initially, a square cuboid with side length $a$ and height $h=a/2$, positioned on a PEC plane is chosen as the antenna structure in the design process, as depicted in Fig.~\ref{fig:cuboidSubduction}(d). However, there is no need to perform a numerical simulation to obtain its CM eigenvalues. Instead, the eigenvalues are simply derived from those of the cube, using the proposed subduction procedure. 

\begin{figure}
    \centering
    \includegraphics[scale=1]{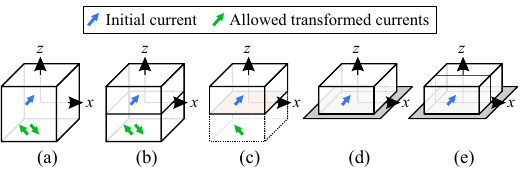}
    \caption{Transition from cube to cuboid on a PEC plane. (a) Cube, $O_\mathrm{h}$ symmetry. (b) Cube sliced in $x$-$y$ plane, $D_\mathrm{4h}$ symmetry. (c) Square cuboid and image, $C_\mathrm{4v}$ symmetry. (d) Square cuboid on PEC plane, $C_\mathrm{4v}$ symmetry. (e) Rectangular cuboid on PEC plane, $C_\mathrm{2v}$ symmetry.}
    \label{fig:cuboidSubduction}
\end{figure}

First, the cube with side length $a$ depicted in Fig.~\ref{fig:cuboidSubduction}(a) ($O_\mathrm{h}$ symmetry) is sliced in the $xy$ plane. {\color{newTextColor}This discrete modification is comparable to the meshing of the spherical shell in Section~\ref{sec:changeSymmetryGroup}, as it only affects electromagnetic behavior through the addition of MICAs.} The resulting structure has $D_\mathrm{4h}$ symmetry, see Fig.~\ref{fig:cuboidSubduction}(b). Its irreps are subduced from those of $O_\mathrm{h}$ using (\ref{eq:projectCurrentOnIrrep}) and given in Table~\ref{tab:C4vSubducedByOh}. 

\begin{table}
\renewcommand{\arraystretch}{1.6}
\caption{$C_\mathrm{4v}$ Irreps That Satisfy (\ref{eq:innerProduct}), Subduced by $O_\mathrm{h}$ irreps: \\How Often Are $C_\mathrm{4v}$ Irreps Contained in $O_\mathrm{h}$ Irreps}
\label{tab:C4vSubducedByOh}
\centering
\begin{tabular}{|c|c|c|c|c||c|}
\hline
\makecell{$O_\mathrm{h}$ \\ irrep} & \makecell{$D_\mathrm{4h}$ \\ irrep} & $\mathbf{\Gamma}^{(p)}_\mathrm{D4h}(IC_\mathrm{2z})$ & \makecell{$D_\mathrm{4h}$ irrep \\ odd only} & \makecell{$C_\mathrm{4v}$ irrep \\ odd only} & \makecell{$C_\mathrm{2v}$ irrep \\ odd only} \\
\hline\hline
${A_\mathrm{1g}}$ & $A_\mathrm{1g}$ & 1 & - & - & -\\
\hline
${A_\mathrm{2g}}$ & $B_\mathrm{1g}$ & 1 & - & - & -\\
\hline
\multirow{2}{*}{${E_\mathrm{g}}$} & $A_\mathrm{1g}$ & 1 & - & - & -\\
\cline{2-6}
 & $B_\mathrm{1g}$ & 1 & - & - & -\\
\hline
\multirow{3}{*}{${T_\mathrm{1g}}$} & $A_\mathrm{2g}$ & 1 & - & - & -\\
\cline{2-6}
 & \multirow{2}{*}{$E_\mathrm{g}$} &  \multirow{2}{*}{$\biggl( \begin{array}{cc} -1 & 0 \\ 0 & -1 \end{array} \biggr)$} &  \multirow{2}{*}{$E_\mathrm{g}$} &  \multirow{2}{*}{$E$} &  $B_1$\\
 \cline{6-6}
  & & & & & $B_2$ \\
\hline
\multirow{3}{*}{${T_\mathrm{2g}}$} & $B_\mathrm{2g}$ & $1$ & - & - & -\\
\cline{2-6}
 & \multirow{2}{*}{$E_\mathrm{g}$} &  \multirow{2}{*}{$\biggl( \begin{array}{cc} -1 & 0 \\ 0 & -1 \end{array} \biggr)$} &  \multirow{2}{*}{$E_\mathrm{g}$} &  \multirow{2}{*}{$E$} &  $B_1$\\
 \cline{6-6}
  & & & & & $B_2$ \\
\hline
${A_\mathrm{1u}}$ & $A_\mathrm{1u}$ & $-1$ & $A_\mathrm{1u}$ & $A_2$ & $A_2$\\
\hline
${A_\mathrm{2u}}$ & $B_\mathrm{1u}$ & $-1$ & $A_\mathrm{2u}$ & $B_2$ & $A_2$\\
\hline
\multirow{2}{*}{${E_\mathrm{u}}$} & $A_\mathrm{1u}$ & $-1$ & $A_\mathrm{1u}$ & $A_2$ & $A_2$\\
\cline{2-6}
 & $B_\mathrm{1u}$ & $-1$ & $B_\mathrm{1u}$ & $B_2$ & $A_2$\\
\hline
\multirow{3}{*}{${T_\mathrm{1u}}$} & $A_\mathrm{2u}$ & $-1$ & $A_\mathrm{2u}$ & $A_1$ & $A_1$\\
\cline{2-6}
 & $E_\mathrm{u}$ & $\biggl( \begin{array}{cc} 1 & 0 \\ 0 & 1 \end{array} \biggr)$ & - & - & -\\
\hline
\multirow{3}{*}{${T_\mathrm{2u}}$} & $B_\mathrm{2u}$ & $-1$ & $B_\mathrm{2u}$ & $B_1$ & $A_1$\\
\cline{2-6}
 & $E_\mathrm{u}$ & $\biggl( \begin{array}{cc} 1 & 0 \\ 0 & 1 \end{array} \biggr)$ & - & - & -\\
\hline
\end{tabular}
\end{table}

Second, the effect of the PEC plane on the modal current distributions is introduced. 
From image theory, it is known that the addition of a PEC plane is electromagnetically equivalent to a mirroring of the currents from the top half-space at the PEC plane with an additional 180° phase shift \cite{Balanis2016}. 
Consequently, only those modes of the cube that fulfill this criterion are also modes of the cuboid on the PEC plane. This limitation to odd modes is indicated by the blue and green arrows in Fig.~\ref{fig:cuboidSubduction}. 
The mirroring at the $xy$ plane is described by the symmetry operation $IC_\mathrm{2z}$. For an odd mode that is a basis function $\boldsymbol\psi^{(p)}$ of the $p$th irrep, 
\begin{equation} \label{eq:innerProduct}
    \langle \boldsymbol\psi^{(p)}, P(IC_\mathrm{2z}) \boldsymbol\psi^{(p)} \rangle = -1
\end{equation}
must be satisfied, assuming $\boldsymbol\psi^{(p)}$ is normalized. This is easily verified by mirroring the blue arrow in Fig.~\ref{fig:cuboidSubduction}(c) at the $xy$ plane, which yields the green arrow with a 180° phase shift, as desired. 
Fortunately, evaluating (\ref{eq:innerProduct}) is not necessary since it is equivalent to the definition of the irrep matrices (\ref{eq:irrepDefinition}) with $\mu=\nu$. These are found in the literature \cite{Cornwell1997} and given in Table~\ref{tab:C4vSubducedByOh} as $\mathbf{\Gamma}^{(p)}_\mathrm{D4h}(IC_\mathrm{2z})$.
The irreps that satisfy $\Gamma^{(p)}_{\mu\mu}(IC_\mathrm{2z}) = -1$, and thereby possess only odd basis functions, are provided in the third column of Table~\ref{tab:C4vSubducedByOh}. 

Finally, the symmetry group of the cuboid antenna structure with height $a/2$ on a PEC plane depicted in Fig.~\ref{fig:cuboidSubduction}(d) is the $C_\mathrm{4v}$ group. It has the 1-D irreps $A_1$, $B_1$, $A_2$ and $B_2$, and the 2-D irrep $E$. These irreps are subduced from the remaining ones of $D_\mathrm{4h}$ according to the fourth and fifth columns of Table~\ref{tab:C4vSubducedByOh}. In this case, the subduction is equivalent to a pairwise mapping between the irreps. Note that it is also possible to enforce (\ref{eq:innerProduct}) on $O_\mathrm{h}$ directly and subduce $C_\mathrm{4v}$ from the result. However, using $D_\mathrm{4h}$ as an intermediate step provides a clearer interpretation. Also note that the validity of (\ref{eq:innerProduct}) is not limited to this specific example (see also \cite{Schab2017}). It can be utilized to assess the influence of a PEC plane on {\color{modTextColor}antenna structures that are symmetric with respect to this plane, but are otherwise arbitrary. Additionally, \mbox{2-D} structures can be investigated} (see Appendix~\ref{sec:cubeToPlanar}).
{\color{newTextColor}Lastly, note that applying (\ref{eq:innerProduct}) to some groups can lead to (\ref{eq:subduction}) yielding noninteger values. This happens if the dimension of an irrep matrix is reduced, but is not set to zero. In this case, $n_p$ contains a relative metric for how many orthogonal basis functions remain. }

\subsection{Insight Gained Through Subduction}
\label{sec:insightInExample}
Table~\ref{tab:C4vSubducedByOh} not only provides all information required to derive the CM eigenvalue traces of the initial antenna structure from those of the cube. It also yields a concise overview of its key properties. In particular, it shows that, e.\,g., $T_\mathrm{1g}$ and $T_\mathrm{2g}$ modes are now basis functions of the same irrep $E$. Therefore, a port designed for the subduced irrep can now be correlated with modes from both original irreps and its far-field will be a superposition of the far-fields of these modes \cite{Peitzmeier2019}. }

\begin{figure}
    \centering
    \includegraphics[scale=0.5]{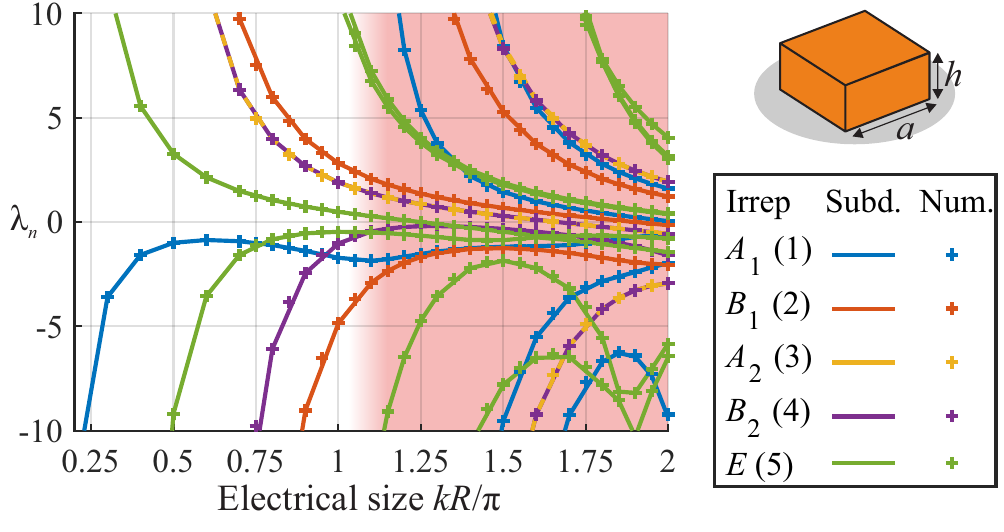}
    \caption{Eigenvalue traces of selected CMs of a square cuboid with height $h=a/2$, positioned on a PEC plane. Colors reveal the assignment to the irreps of the $C_\mathrm{{4v}}$ group, given in the legend. Solid lines show the prediction obtained through subduction from the cube, and markers provide direct simulation results for reference. The predicted frequency region with undesired far-field patterns is indicated in pink.}
    \label{fig:halfCubeOnGND}
\end{figure}

{The eigenvalue traces resulting from this subduction procedure for the initial antenna structure are depicted in Fig.~\ref{fig:halfCubeOnGND}. Reference results obtained through a numerical simulation of the antenna structure on the PEC plane prove that the proposed procedure accurately predicts eigenvalues and irreps. 
Note that to determine the electrical size of the structure, the center of the circumscribing sphere remains inside the PEC plane. This way, the image created by mirroring the structure at the plane is included and results remain comparable to those of the cube. 

Now, these results are compared to those expected for a cuboid with rectangular footprint, see Fig.~\ref{fig:cuboidSubduction}(e). The last column of Table~\ref{tab:C4vSubducedByOh} contains the subduced irreps of the $C_\mathrm{2v}$ group, which is the symmetry group of a rectangular cuboid on a PEC plane and possesses four 1-D irreps. Crucially, both the $T_\mathrm{1u}$ and the $T_\mathrm{2u}$ irrep of $O_\mathrm{h}$ subduce the same irrep in $C_\mathrm{2v}$ ($A_1$), while they subduce two different irreps in $C_\mathrm{4v}$ ($A_1$ and $B_1$). Consequently, the blue and red traces in Fig.~\ref{fig:halfCubeOnGND} would belong to the same irrep in the former case. As is clarified in the following, a port pattern similar to that of the first $A_1$ mode of the square cuboid is desirable for the application. The presence of additional modes in this irrep makes it more difficult to achieve this goal since the correlation of a greater variety of modes to an irrep's port needs to be considered. Therefore, the square cuboid is selected as the preferable structure, based on the findings of the subduction procedure.

For the investigated direction-finding application, at least three independent port far-fields are required. A set consisting of a monopole-type mode and two orthogonal loop modes (magnetic dipoles) in the $xz$ and $yz$ planes is shown to be a well-suited choice in \cite{Grundmann2023}. Using Tables~\ref{tab:OhSubducedByO3} and \ref{tab:C4vSubducedByOh}, it is easy to identify these modes in Fig.~\ref{fig:halfCubeOnGND}, without the need to numerically calculate the far-fields: It is known from the literature that the modes of the $\Gamma_\mathrm{O(3)}^\mathrm{(1)TM}$ irrep are of the electric dipole type, while those associated with $\Gamma_\mathrm{O(3)}^\mathrm{(1)TE}$ belong to the magnetic dipole type \cite{Hansen1988}. 
These irreps subduce $T_\mathrm{1u}$ and $T_\mathrm{1g}$ in the $O_\mathrm{h}$ symmetry group, which in turn subduce $A_1$ and $E$ in $C_\mathrm{4v}$, respectively. Here, the PEC plane causes the far-fields to be zero below the $xy$ plane and even modes are ruled out. Consequently, the electric monopole mode is the first blue trace in Fig.~\ref{fig:halfCubeOnGND}, while the two degenerate loop modes correspond to the first green trace with positive eigenvalue.

However, the first mode from the $A_1$ irrep contains a crossing avoidance near $kR/\pi\approx1.1$. Consequently, the mode no longer possesses the desired pure monopole-type far-field pattern, around and above this electrical size, which is indicated in pink in Fig.~\ref{fig:halfCubeOnGND}. The presence of the MACA is identified either by following its origins through Fig.~\ref{fig:sph2cube} and Table~\ref{tab:C4vSubducedByOh}, or through the characteristic indentation in the eigenvalue trace in Fig.~\ref{fig:halfCubeOnGND}. Since the second mode associated with the MACA is no longer near significance, the indentation in the CM trace will directly influence the matching and far-field of the port that is to be designed in the following.


So far, without ever performing a numerical simulation of the actual structure under investigation, we were able to identify key features of its CMs and possible ports. In addition to the simulation effort already saved, this insight provides design guidance and serves as a sanity check for simulation results in the following numerical optimization.}

\begin{figure}
    \centering
    \includegraphics[scale=0.5]{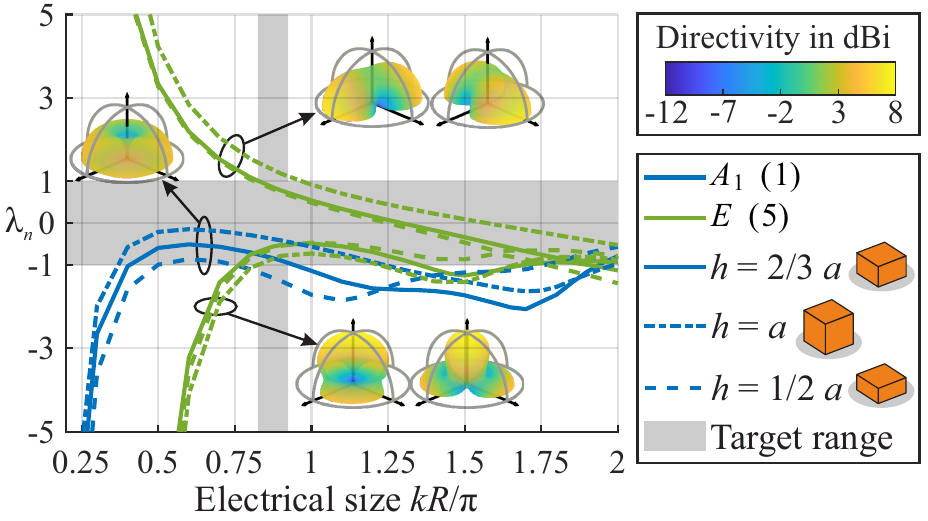}
    \caption{Eigenvalue traces of relevant CMs for a cuboid with variable height on a PEC plane. The cuboid with height $h=2a/3$ is selected as the fundamental structure of the demonstrator antenna system. The far-fields for $kR/\pi\approx 0.87$ are depicted for this case.}
    \label{fig:EVantenna}
\end{figure}

{\subsection{Implementation}
\label{sec:implementation}
Now, numerical simulations are performed to tune the eigenvalue traces of the desired modes toward modal significance ($|\lambda_n|<1$).
{\color{newTextColor}These simulations are required since the subduction procedure cannot provide exact results for arbitrarily parameterized variations of a structure, compare Fig.~\ref{fig:sph2cube}.} 
Eigenvalue traces for three different cuboid heights $h$ are depicted in Fig.~\ref{fig:EVantenna}, together with the typical modal far-fields at $kR/\pi\approx 0.87$. 
The antenna shall have minimal electrical size and the shape of the modal far-fields shall have the desired characteristic across the target frequency range.} 

Starting with the height $h=a/2$ (dashed lines), it is found that the desired modes of the $E$ irrep reach $\lambda_n=1$ at $kR/\pi \approx 0.82$. Increasing the height to $h=a$ (dashed-dotted lines) increases the electrical size required to ${kR/\pi \approx 0.97}$. 
The monopole mode from the $A_1$ irrep is significant in the range $0.37 \leq kR/\pi \leq 1.22$ in this case. However, reducing the height to $h=a/2$ shifts its range of significance to smaller electrical sizes. Due to the indentation in the eigenvalue trace of the monopole mode ($A_1$ irrep), there is no common range of $kR/\pi$ in which all desired modes are significant for $h=a/2$. 
Consequently, a compromise is chosen in $h=2a/3$. With this setup, all desired modes are significant for ${0.82 \leq kR/\pi \leq 0.92}$, which is marked in Fig.~\ref{fig:EVantenna}. {For a circumscribing sphere radius of $R=123.7\,\mathrm{mm}$, this leads to a frequency range between 1\,GHz and 1.12\,GHz. This includes the target frequency range and a margin, while the MACA is avoided and thereby the intended far-field characteristic is maintained.} 

{On the downside, it is found that two undesired modes from the $E$ irrep are also significant in this frequency range, as predicted (bottom green traces in Fig.~\ref{fig:EVantenna}).} 
The significance of the desired modes could be increased by increasing the electrical size of the antenna. Simultaneously, this would reduce the impact of the undesired modes. However, as predicted above, the MACA of the $A_1$ mode near $kR/\pi = 1.1$ leads to an indentation in the eigenvalue trace. {Importantly, it is known from the prediction in Section~\ref{sec:insightInExample} that the shape of the far-field pattern also changes due to the crossing avoidance.} This effect negates any advantages gained through the higher significance of the desired $E$ modes. {Therefore, the presence of the undesired modes is accepted as a compromise. Overall, the characteristics of these simulation results match the behavior predicted above.}

\begin{figure}
    \centering
    \subfloat[(a)]{
    \includegraphics[]{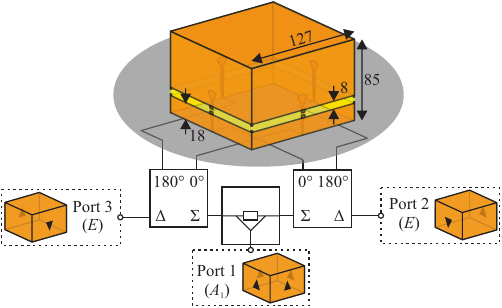}
    \label{fig:antennaCAD:sim}}
    \hfil
    \subfloat[(b)]{
    \includegraphics[]{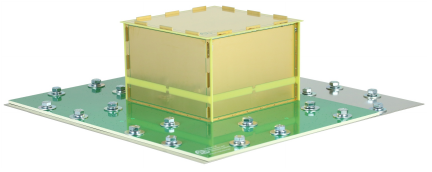}
    \label{fig:antennaCAD:real}}
    \caption{\protect\subref{fig:antennaCAD:sim} Idealized semitransparent simulation model and \protect\subref{fig:antennaCAD:real} photograph of the demonstrator. The antenna is first presented in \cite{Grundmann2023}. The cuboid antenna structure with $h=2a/3$ is positioned on a PEC plane and connected to the ports via a feed network. Dimensions are given in mm.}
    \label{fig:antennaCAD}
\end{figure}

{ With the outer dimensions of the antenna decided on, we briefly recapitulate the development of the ports that excite the desired far-fields from \cite{Grundmann2023}, following the procedure from \cite{Peitzmeier2022}. It starts with a current vector positioned at the center of one of the lateral surfaces of the cuboid and directed perpendicular to the PEC plane, which we call a feed. 
By applying (\ref{eq:projectCurrentOnIrrep}) to it, the port setups depicted in the dashed boxes in Fig.~\subref*{fig:antennaCAD:sim} are created as superpositions of discrete feeds. Since these are respective basis functions of the $A_1$ and $E$ irreps, they exclusively correlate with CMs from the respective irrep. 
For further details, refer to \cite{Grundmann2023}, where design aspects of this antenna related to its direction-finding capabilities are provided, as introduced in Section~\ref{sec:Introduction}.
Since each port is connected to multiple feeds, the signals from the ports need to be distributed using a feed network. A feed network consisting of two hybrid couplers and a Wilkinson power divider is employed in Fig.~\subref*{fig:antennaCAD:sim}. 
For each of the four feeds, a microstrip line leads up the inside of the respective surface of the cuboid. The signal is coupled to the outer surface of the antenna structure through a slot that spans the circumference of the lateral surfaces. 
Numerical optimizations are conducted to determine the dimensions of the microstrip lines and the slot. Notches are added to the center and the ends of the slot on each lateral surface to further tune the input reflection coefficients of the ports. The resulting simulation model is depicted in Fig.~\subref*{fig:antennaCAD:sim}. 
From this, a demonstrator is manufactured as depicted in Fig.~\subref*{fig:antennaCAD:real}. The surface of the manufactured antenna is made of five individual printed circuit boards (PCBs), while the feed network is integrated into the green PCB positioned below the cuboid \cite{Grundmann2023}.}

\begin{figure}
    \centering
    \includegraphics[scale=0.5]{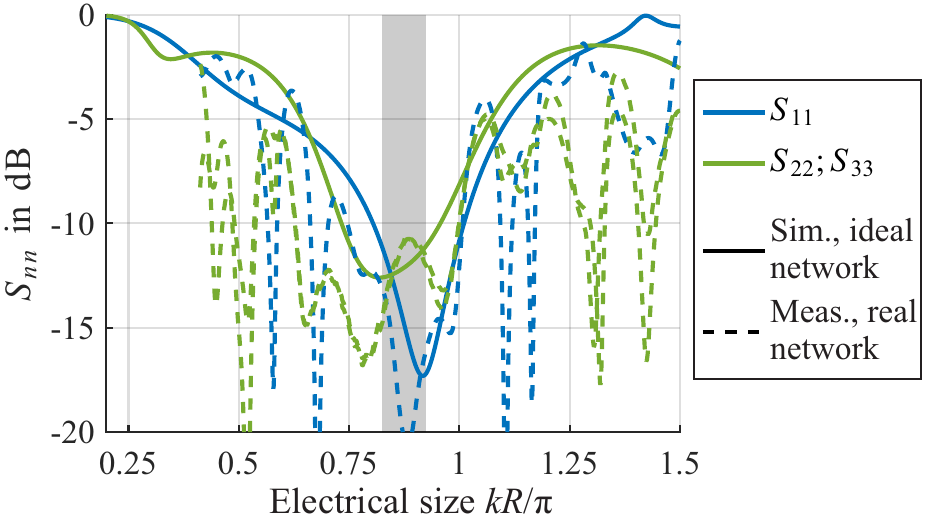}
    \caption{Simulated and measured input reflection coefficients of the demonstrator antenna from Fig.~\ref{fig:antennaCAD}. Solid lines represent results from a finite-difference time-domain (FDTD) simulation of the antenna, where the feed network consists of ideal couplers, see Fig.~\protect\subref*{fig:antennaCAD:sim}. Dashed lines show measurement results for the demonstrator depicted in Fig.~\protect\subref*{fig:antennaCAD:real}. }
    \label{fig:antennaMatching}
\end{figure}
The simulated input matching of the three ports, using the ideal feed network, is represented by the solid lines in Fig.~\ref{fig:antennaMatching}. An input reflection coefficient below $-10\,\mathrm{dB}$ is achieved for all ports across the desired frequency range. This observation also holds for a measurement of the antenna with the realized feed network (dashed lines). Here, peaks and minima are due to the limited bandwidth of the utilized couplers. 

\section{Conclusion} \label{sec:conclusion}
The presented procedure promotes the understanding of CM eigenvalue behavior based on group theory. This provides additional insight for antenna design, as CMs are an established tool for understanding antennas. {An analytical connection between well-understood solutions for structures of higher symmetry and CMs of lower symmetry structures is found in the subduction of irreps. In combination with numerical analysis, this systematically provides interpretations and predictions for the behavior of many practically relevant antenna structures.} One example of an interpretation obtained from this is the MACA. It provides an explanation for the indentations in the eigenvalue traces observed for the increasingly important 3-D antenna structures. 
The practical relevance of this observation is exemplified by the demonstrator antenna design. {Due to the predictions made, parameters are chosen that simplify input matching and improve far-field stability over frequency.} 

Prediction of MACA is also relevant for CM tracking, where sharp peaks in eigenvalue traces are particularly challenging \cite{Masek2020}. This can be addressed by increasing the frequency resolution in areas where MACAs are expected. {Also, subduction is a perquisite for tracking CMs across different geometries with varying symmetry.}


%

\appendices
{
\section{Transition to Planar Structures} \label{sec:cubeToPlanar}
\begin{figure}
    \centering
    \subfloat[(a)]{
    \includegraphics[scale=0.5]{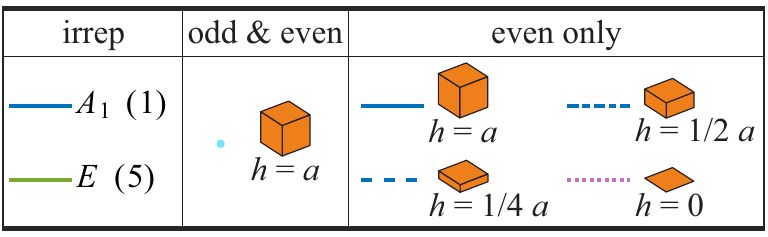}
    \label{fig:cube2planar:leg}}
    \hfil
    \subfloat[(b)]{
    \includegraphics[width=0.45\columnwidth]{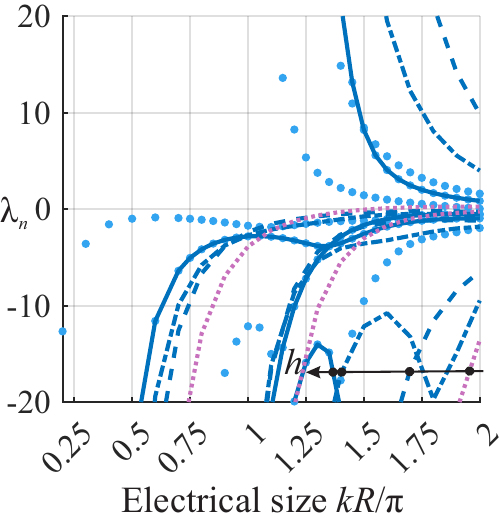}
    \hfil
    \label{fig:cube2planar:irrep1}}
    \subfloat[(c)]{
    \includegraphics[width=0.45\columnwidth]{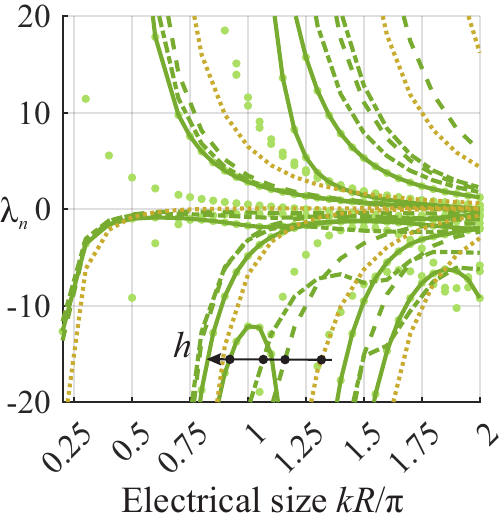}
    \label{fig:cube2planar:irrep5}}
    \caption{Selection of eigenvalue traces for different heights (parameter $h$) of the square cuboid in free-space, see \protect\subref{fig:cube2planar:leg}. The geometries are identified by line style. \protect\subref{fig:cube2planar:irrep1} and \protect\subref{fig:cube2planar:irrep5} contain selected CM eigenvalue traces belonging to the $C_\mathrm{4v}$ irreps $A_1$ and $E$, respectively. Results for a square plate in free-space are highlighted in a different color. Black arrows indicate the impact of changes to $h$ on the CM eigenvalue traces marked by black dots.}
    \label{fig:cube2planar}
\end{figure}

The transition of the eigenvalue traces on a cube ($O_\mathrm{h}$) to those on a square plate ($C_\mathrm{4v}$) with height $h=0$ in free-space is illustrated for a selection of modes in Fig.~\ref{fig:cube2planar}. Only modes that are even with respect to the $xy$ plane can exist on the plate, which means $\Gamma^{(p)}_{\mu\mu}(IC_\mathrm{2z}) = 1$, compare (\ref{eq:innerProduct}).
Consequently, a subduction procedure in analogy to Section~\ref{sec:subductionToC4v} is applied to the set of all modes on the cube (dots in Fig.~\ref{fig:cube2planar}). The remaining modes are even (solid traces in Fig.~\ref{fig:cube2planar}).
Similarly, the subduction from $D_\mathrm{4h}$ to $C_\mathrm{4v}$ is applied to the numerically calculated CM eigenvalues of two cuboids with decreasing height, using only even modes.

Due to the limitation to even modes, many crossing avoidances observed before are ruled out. For example, the leftmost trace of dots in Fig.~\subref*{fig:cube2planar:irrep1} has no corresponding solid trace. Interestingly, no additional crossing avoidances are introduced by the subduction in the depicted frequency range although this observation does not hold if the frequency range is extended. 
Moreover, the remaining MACAs are shifted to higher frequencies and become less pronounced with decreasing $h$, which is observable for the modes marked by the black arrow in Fig.~\ref{fig:cube2planar}.

In combination, these effects explain the absence of observable eigenvalue trace indentations for many \mbox{2-D} structures, see Section~\ref{sec:Introduction} and Fig.~\subref*{fig:MotivationEVPlate}. Moreover, a link between the CMs of the plate and the SWFs is established. Note that these effects reduce the likelihood of observable MACAs, but do not fundamentally prohibit them. Therefore, MACAs can still appear for higher frequencies or if the structure is modified, for example in appendix~\ref{sec:otherStructures}.}

\section{Dipole-like wire structures} \label{sec:wireStructures}
\begin{figure}
    \centering
    \includegraphics[scale=0.5]{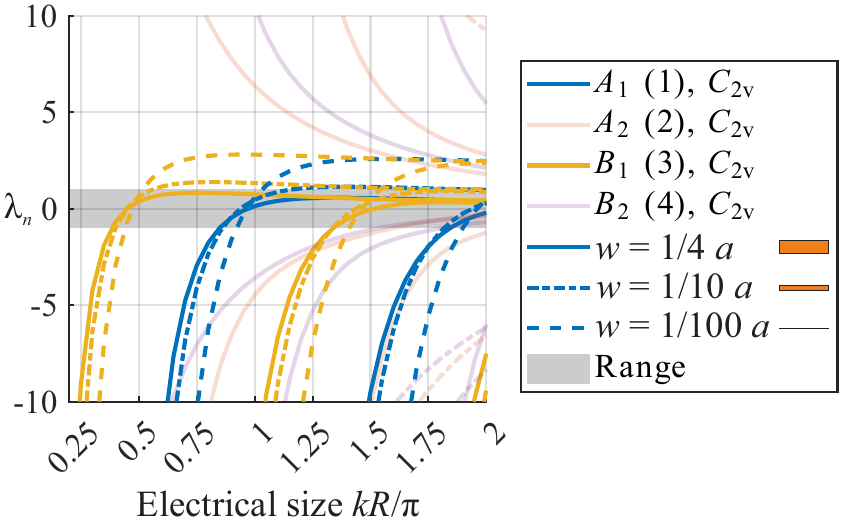}
    \caption{Eigenvalues of a dipole with length $a$ and varying width $w$, indicated by different line styles. The symmetry group is $C_\mathrm{2v}$.}
    \label{fig:EVdipole}
\end{figure}
Planar wire structures are among the most common antenna structures as they are used for dipole- and monopole-type antennas. We provide a brief clarification with respect to MACAs, to avoid possible misconceptions for this type of antenna. 

A planar dipole is of the $C_\mathrm{2v}$ symmetry group, which possesses four 1-D irreps. These are given in the legend of Fig.~\ref{fig:EVdipole}. 
The CM eigenvalues for a planar dipole of varying width $w$ are shown in Fig.~\ref{fig:EVdipole}. It is found that for smaller widths, the eigenvalue traces of the dipole modes are shifted toward higher eigenvalues, see the yellow and blue traces in Fig.~\ref{fig:EVdipole}. This observation is not to be mistaken for a MACA, with the hypothetical second trace above the displayed eigenvalue range. An indicator for this is that the direction of the trace's curvature does not change. Consequently, the shape of the far-field pattern is also maintained. An example is the first, yellow, dashed trace in the range $kR/\pi<1.5$. 

\section{Structures With Holes} \label{sec:otherStructures}
\begin{figure}
    \centering
    \includegraphics[scale=0.5]{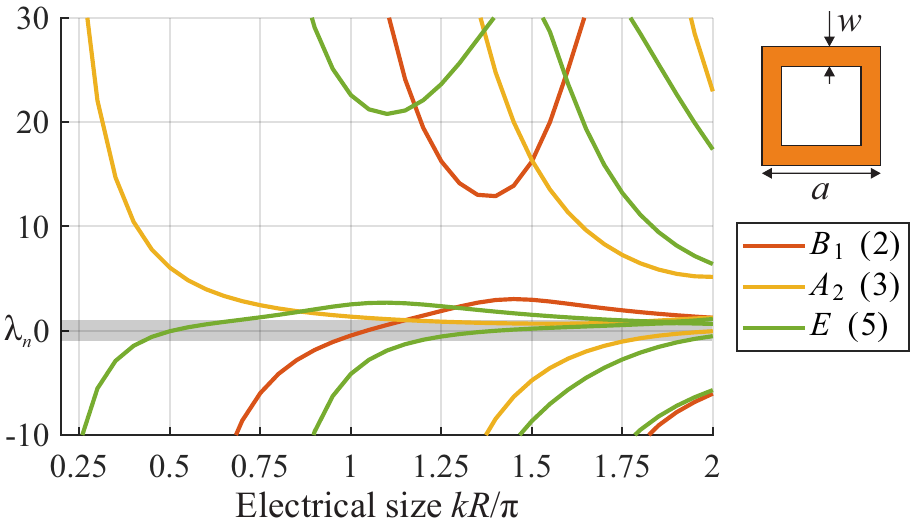}
    \caption{Eigenvalues of a square loop structure with width $w=a/6$, where $a$ is the side length of the outer square. This is an example with eigenvalue trace indentations towards positive infinity and corresponding peaks. Only a selection of traces is shown for simplicity.}
    \label{fig:EVloop}
\end{figure}
The antenna structures investigated in this work are all convex and do not possess holes. This approach is apparent since the spherical shell also fulfills these conditions. {The poles in the eigenvalue traces of the spherical shell go from negative infinity to positive infinity, and never in the opposite direction. Therefore, the observed indentations only ever appear toward negative eigenvalues, never toward positive ones. However, this observation does not hold for structures with holes.} 

Fig.~\ref{fig:EVloop} shows a selection of eigenvalue traces for a structure that contains a hole. Thereby, no continuous mapping of this topology to the spherical shell is possible. The example structure is a square loop of $C_\mathrm{4v}$ symmetry. Fig.~\ref{fig:EVloop} reveals a MACA near $kR/\pi\approx 1.1$ for modes from the $E$ irrep, and a second one near $kR/\pi\approx 1.3$ for modes from the $B_1$ irrep. The eigenvalue trace indentations are toward positive infinity in this case, in contrast to the MACAs observed {above}. 
A similar behavior is observed for 3-D toroidal structures. These are not depicted here to avoid the introduction of an additional symmetry group.
{Nonetheless, this finding underlines that the proposed method yields interpretations that are adaptable to different categories of structures as well.} 

\ifCLASSOPTIONcaptionsoff
  \newpage
\fi



\bibliographystyle{IEEEtran}
\bibliography{IEEEabrv,literatur_master360}

\begin{IEEEbiography}[{\includegraphics[width=1in,height=1.25in,clip,keepaspectratio]{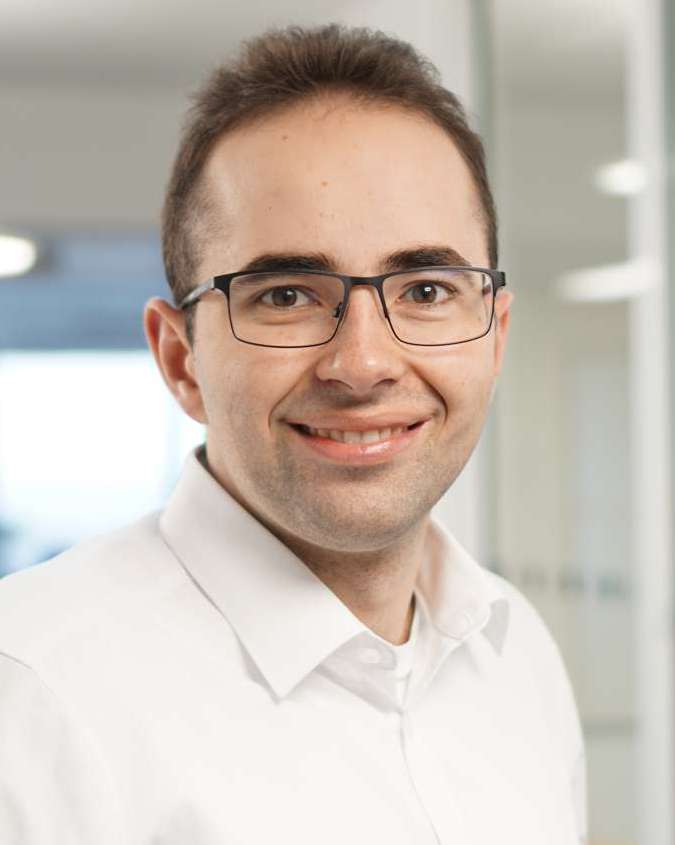}}]{Lukas Grundmann} (Graduate Student Member, IEEE) was born in Hoya, Germany, in 1994. He received the B.Sc. and M.Sc. degrees in electrical engineering from Leibniz University Hannover, Hannover, Germany, in 2017 and 2019, respectively. He is currently a Research Assistant with the Institute of Microwave and Wireless Systems, Leibniz University Hannover. His current research focuses on modal expansion techniques, such as spherical wave functions and characteristic modes, and their applications to antenna development. 
\end{IEEEbiography}

\begin{IEEEbiography}[{\includegraphics[width=1in,height=1.25in,clip,keepaspectratio]{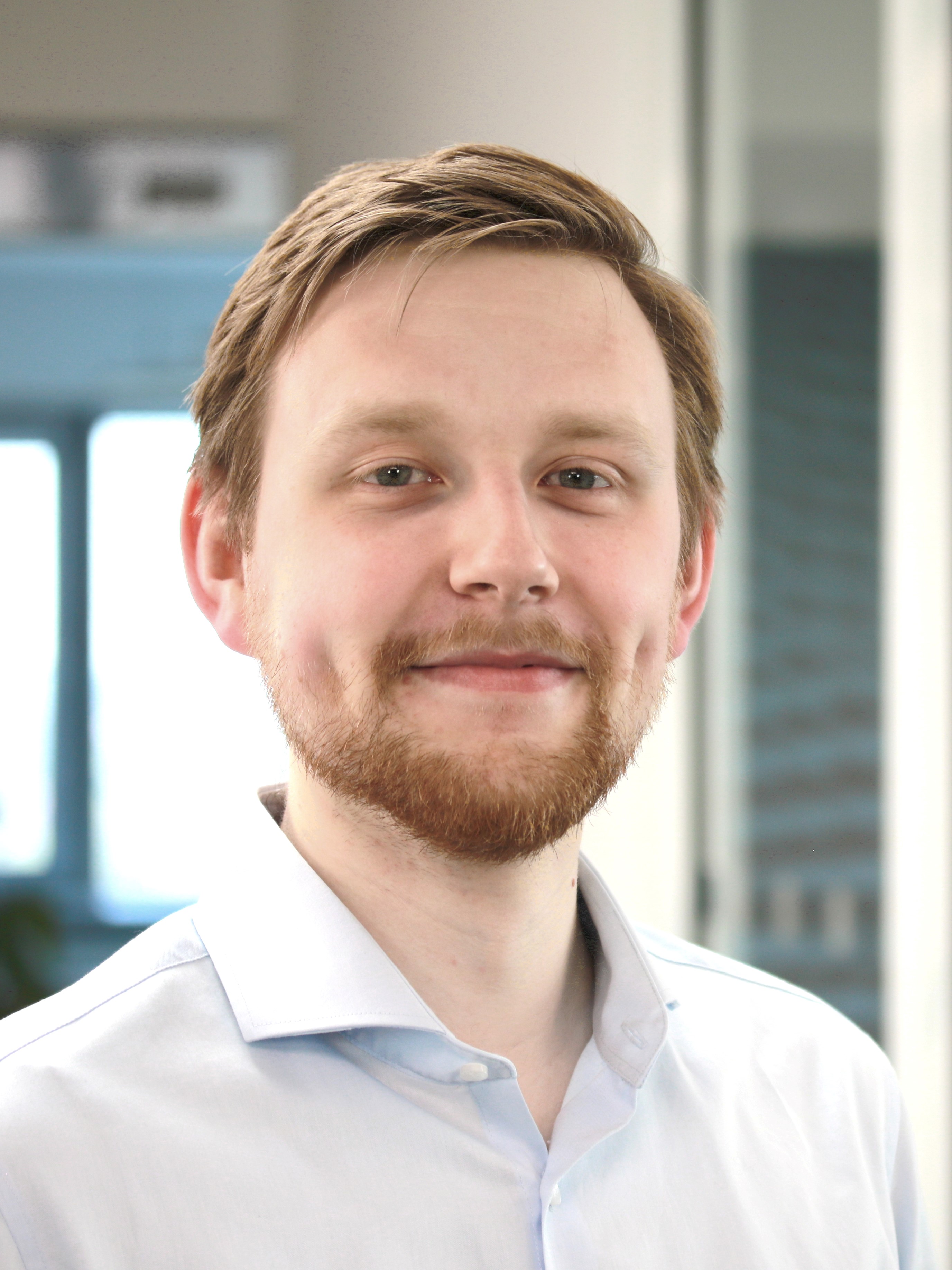}}]{Lukas Warkentin} was born in Elmshorn, Germany, in 1999. He received the B.Sc. in electrical engineering from Hamburg University of Applied Sciences, Hamburg, Germany in 2021 and the M.Sc. in electrical engineering from Leibniz University Hannover, Hannover, Germany, in 2023. He is currently a Research Assistant with the Institute of Microwave and Wireless Systems, Leibniz University Hannover. His current research interest include the theory of characteristic modes and its connection to group theory.
\end{IEEEbiography}

\begin{IEEEbiography}[{\includegraphics[width=1in,height=1.25in,clip,keepaspectratio]{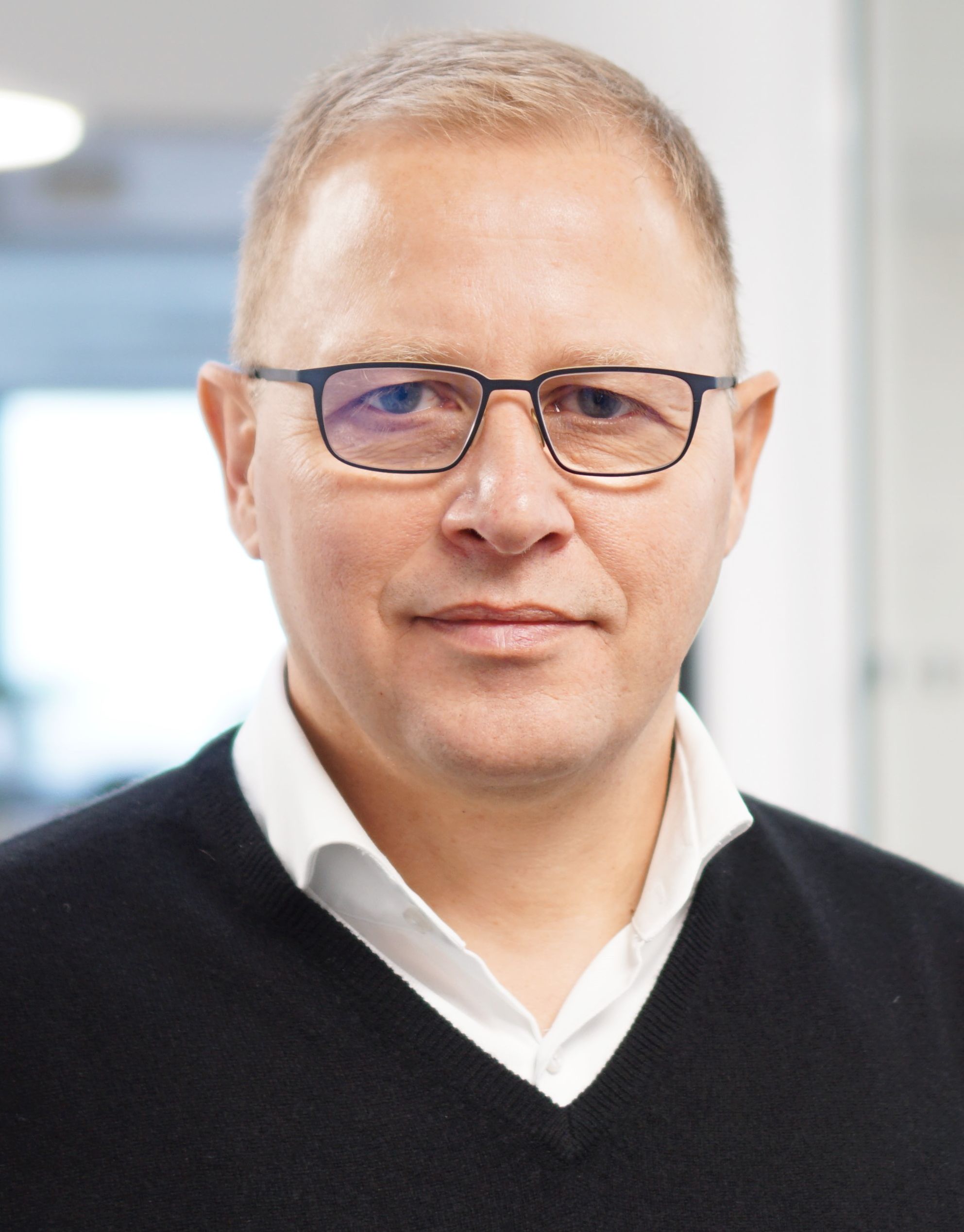}}]{Dirk Manteuffel} (Member, IEEE) was born in Issum, Germany, in 1970. He received the Dipl.-Ing. and Dr.-Ing. degrees in electrical engineering from
the University of Duisburg–Essen, Duisburg, Germany, in 1998 and 2002, respectively.
From 1998 to 2009, he was with IMST, Kamp-Lintfort, Germany. As a Project Manager, he was responsible for industrial antenna development and advanced projects in the field of antennas and electromagnetic (EM) modeling. From 2009 to 2016, he was a Full Professor of wireless communications at Christian-Albrechts-University, Kiel, Germany. Since June 2016, he has been a Full Professor and the Executive Director of the Institute of Microwave and Wireless Systems, Leibniz University Hannover, Hannover, Germany. His research interests include electromagnetics, antenna integration and EM modeling for mobile communications and biomedical applications.
Dr. Manteuffel was a Director of the European Association on Antennas and Propagation from 2012 to 2015. He served on the Administrative Committee (AdCom) of IEEE Antennas and Propagation Society from 2013 to 2015 and as an Associate Editor of the IEEE Transactions on Antennas and Propagation from 2014 to 2022. Since 2009 he has been an Appointed Member of the Committee "Antennas" of the German VDI-ITG. 

\end{IEEEbiography}

\end{document}